\documentclass[conference]{IEEEtran}
\IEEEoverridecommandlockouts

\usepackage{cite}
\usepackage[utf8]{inputenc}
\usepackage{tgtermes}
\usepackage{balance}
\usepackage{amsmath,amssymb,amsfonts}
\usepackage{amsthm}
\usepackage[table]{xcolor}
\usepackage{colortbl}
\usepackage{soul}
\usepackage[english]{babel}
\newtheorem{theorem}{Theorem}
\usepackage{graphicx}
\usepackage{wrapfig}
\usepackage{subcaption}
\usepackage{tikz}
\usetikzlibrary{automata,arrows.meta,positioning,shapes}
\usepackage{pgfplots}
\usepackage{pgfplotstable}
\pgfplotsset{compat=1.9}
\usepackage{pgf-pie}

\usepackage{tabularx}
\usepackage{tabularray}
\usepackage{threeparttable}
\usepackage{booktabs}
\usepackage{multirow}
\usepackage{multicol}
\usepackage{indentfirst}
\usepackage{balance}

\usepackage{tikz}
\usepackage{pgfplots}
\pgfplotsset{compat=1.18}

\usepackage{algorithm}
\usepackage{algorithmicx}
\usepackage{algpseudocode}
\usepackage{listings}
\usepackage{listingsutf8}
\lstdefinelanguage{json}{
  basicstyle=\ttfamily\footnotesize,
  breaklines=true,
  keywordstyle=\color{red}\bfseries,  
  stringstyle=\color{black},
  commentstyle=\color{gray},
  showstringspaces=false,
  numbers=none,
  frame=single,
  framerule=0.5pt,
  rulecolor=\color{black},
  moredelim=[is][\color{red!70!black}\bfseries]{(*@}{@*)},
  backgroundcolor=\color{lightgray!50},
  linewidth=0.98\columnwidth,
}

\lstdefinelanguage{cypher}{
  basicstyle=\ttfamily\footnotesize,
  breaklines=true,
  keywordstyle=\color{green}\bfseries,  
  stringstyle=\color{black},
  commentstyle=\color{gray},
  showstringspaces=false,
  numbers=none,
  frame=single,
  framerule=0.5pt,
  rulecolor=\color{black},
  moredelim=[is][\color{gray!50!black}\bfseries]{(*@}{@*)},
  backgroundcolor=\color{lightgray!50},
  linewidth=0.98\columnwidth,
}

\newcommand{\mylegenditem}[3]{%
    \begin{tikzpicture}[baseline=(legendtext.base)]
        \node(legendbox)[state, shape=rectangle, draw=#1, fill=#2!30, minimum width=4mm, minimum height=4mm, inner sep=1pt] {};
        \node(legendtext)[right=0.5mm of legendbox, anchor=west] {#3};
    \end{tikzpicture}%
}

\usepackage{textcomp}
\usepackage{enumitem}
\usepackage{comment}
\newlist{todolist}{itemize}{2}
\setlist[todolist]{label=$\square$}

\def\BibTeX{{\rm B\kern-.05em{\sc i\kern-.025em b}\kern-.08em
    T\kern-.1667em\lower.7ex\hbox{E}\kern-.125emX}}

\AtBeginDocument{%
  \providecommand\BibTeX{{%
    Bib\TeX}}}

\newtheorem{prop}{Proposition}

\newcommand{\vdgraph}{\texttt{VDGraph}}

\begin{document}

\title{VDGraph: A Graph-Theoretic Approach to Unlock Insights from SBOM and SCA Data}

\author{\IEEEauthorblockN{Howell Xia}
\IEEEauthorblockA{\textit{Boston University}\\
Boston, MA, USA \\
howellx@bu.edu}
\and
\IEEEauthorblockN{Jonah Gluck}
\IEEEauthorblockA{\textit{Boston University}\\
Boston, MA, USA \\
jonahg@bu.edu}
\and
\IEEEauthorblockN{Şevval Şimşek}
\IEEEauthorblockA{\textit{Boston University}\\
Boston, MA, USA \\
sevvals@bu.edu\\}
\and
\IEEEauthorblockN{David Sastre Medina}
\IEEEauthorblockA{\textit{Red Hat Inc.}\\
Madrid, Spain \\
asastrem@redhat.com}
\and 
\IEEEauthorblockN{David Starobinski}
\IEEEauthorblockA{\textit{Boston University}\\
Boston, MA, USA \\
staro@bu.edu}}

\maketitle

\begin{abstract}
The high complexity of modern software supply chains necessitates tools such as Software Bill of Materials (SBOMs) to manage component dependencies, and Software Composition Analysis (SCA) tools to identify vulnerabilities. While there exists limited integration between SBOMs and SCA tools, a unified view of complex dependency-vulnerability relationships remains elusive. In this paper, we introduce \vdgraph, a novel knowledge graph-based methodology for integrating vulnerability and dependency data into a holistic view. \vdgraph\ consolidates SBOM and SCA outputs into a graph representation of software projects' dependencies and vulnerabilities. We provide a formal description and analysis of the theoretical properties of \vdgraph\ and present solutions to manage possible conflicts between the SBOM and SCA data. We further introduce and evaluate a practical, proof-of-concept implementation of \vdgraph\ using two popular SBOM and SCA tools, namely CycloneDX Maven plugin and Google's OSV-Scanner. We apply \vdgraph\ on 21 popular Java projects. Through the formulation of appropriate queries on the graphs, we uncover the existence of concentrated risk points (i.e., vulnerable components of high severity reachable through numerous dependency paths). We further show that vulnerabilities predominantly emerge at a depth of three dependency levels or higher, indicating that direct or secondary dependencies exhibit lower vulnerability density and tend to be more secure. Thus, \vdgraph\ contributes a graph-theoretic methodology that improves visibility into how vulnerabilities propagate through complex, transitive dependencies. Moreover, our implementation, which combines open SBOM and SCA standards with Neo4j, lays a foundation for scalable and automated analysis across real-world projects.
\end{abstract}


\section{Introduction}\label{intro}

Software supply chains have become increasingly complex, in part due to the spread of third-party software. Third-party software has been widely adopted, as developers can leverage open-source software to make product development faster, cheaper, and more streamlined. As a result, many modern software products rely on a complex network of third-party components. However, this widespread adoption of third-party software also raises challenges to security vulnerabilities~\cite{cobleigh2018identifying}, licensing compliance~\cite{gangadharan2008license}, and long-term maintainability~\cite{mari2003impact}. 

Software Bill of Materials (SBOMs) have emerged as a popular way for developers to track their project's dependencies ~\cite{mirakhorli2024landscapestudyopensource}. A well-built SBOM provides a structured document with a comprehensive list of project components and their relations with each other. SBOMs allow developers to better manage their projects by consolidating relevant dependency information. 

Similarly, Software Composition Analysis (SCA) tools are automated tools that help developers address vulnerabilities of open-source components~\cite{haddad2020open}. SCA tools obtain the dependencies of the components of a project and inform the developer of any associated vulnerabilities that result from them. Generally, all SCA tools work by displaying vulnerable components and their resulting vulnerabilities~\cite{sharma2025understanding}.

\begin{figure*}

\par
\centering 
\mylegenditem{red}{orange}{Root Component}\hspace{1em}
\mylegenditem{cyan}{cyan}{Dependency Component}\hspace{1em}
\mylegenditem{violet}{violet}{Vulnerability}
\par\vspace{1em}

\begin{subfigure}{0.4\textwidth}
\begin{tikzpicture} [node distance = 4cm, on grid, auto]

\node (q0) [state, shape=rectangle,style = {draw = red, fill = orange!30}] {flink-parent};
\node (q1) [state, shape=rectangle,style = {draw = cyan, fill = cyan!30}, right = of q0] {flink-runtime};
\node (q2) [state,shape=rectangle, style = {draw = cyan, fill = cyan!30}, below = 2 cm of q1] {hadoop-common};
\node (q3) [state, shape=rectangle,style = {draw = cyan, fill = cyan!30}, left = of q2] {jersey-json};
\node (q4) [state, shape=rectangle,style = {draw = cyan, fill = cyan!30}, below = 2 cm of q3] {jettison};

\path [-stealth, thick]
    (q0) edge node {depn.}(q1)
    (q0) edge node {depn.} (q2)
    (q1) edge node {depn.}(q2)
    (q2) edge node {depn.}(q3)
    (q3) edge node {depn.}(q4);

\end{tikzpicture}
\caption{Only SBOM: No vulnerability information}
\end{subfigure}%
\hfill
\begin{subfigure}{0.2\textwidth}
\begin{tikzpicture} [node distance = 3cm, on grid, auto]
 
\node (q0) [state, shape=rectangle,style = {draw = red, fill = orange!30}] {flink-parent};
\node (q1) [state, shape=rectangle,style = {draw = cyan, fill = cyan!30}, below = 2 cm of q0] {jettison};
\node (q2) [state, shape=rectangle,style = {draw = violet, fill = violet!30}, below = 2 cm of q1] {CVE-2022-45685};

\path [-stealth, thick]
    (q0) edge node {depn.}(q1)
    (q1) edge node {has\_v.}(q2);
    
\end{tikzpicture}
\caption{Only SCA: No component relation data}
\end{subfigure}%
\hfill
\begin{subfigure}{0.4\textwidth}

\begin{tikzpicture} [node distance = 3.5cm, on grid, auto]
 
\node (q0) [state, shape=rectangle,style = {draw = red, fill = orange!30}] {flink-parent};
\node (q1) [state, shape=rectangle,style = {draw = cyan, fill = cyan!30}, right = 4 cm of q0] {flink-runtime};
\node (q2) [state,shape=rectangle, style = {draw = cyan, fill = cyan!30}, below = 2 cm of q1] {hadoop-common};
\node (q3) [state, shape=rectangle,style = {draw = cyan, fill = cyan!30}, left = 4 cm of q2] {jersey-json};
\node (q4) [state, shape=rectangle,style = {draw = cyan, fill = cyan!30}, below = 2 cm of q3] {jettison};
\node (q5) [state, shape=rectangle,style = {draw = violet, fill = violet!30}, right = 4 cm of q4] {CVE-2022-45685};

\path [-stealth, thick]
    (q0) edge node  {depn.}(q1)
    (q1) edge node  {depn.}(q2)
    (q0) edge node {depn.}(q2)
    (q2) edge node {depn.}(q3)
    (q3) edge node  {depn.}(q4)
    (q4) edge node  {has\_v.}(q5);
\end{tikzpicture}
\caption{SBOM+SCA: Full dependency chain}
\end{subfigure}

\caption{\vdgraph\ combines an SBOM subgraph with an SCA subgraph to create a full dependency chain from the root to each vulnerability (the figure shows an example for one vulnerability).}
\label{dependency-chain-image}
\end{figure*}
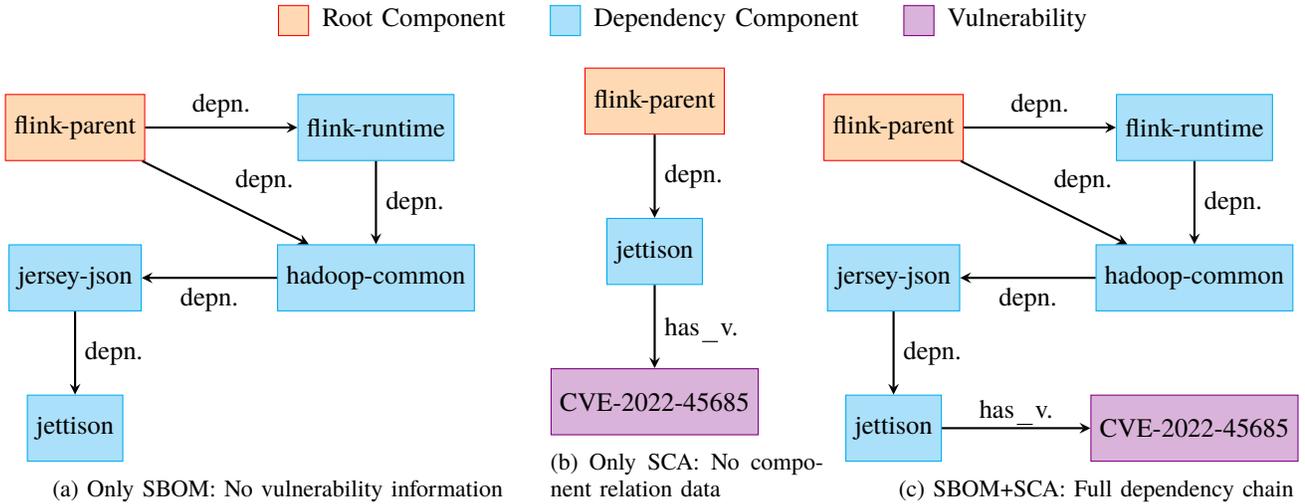

Both of these software management tools help developers track their usage of third-party software but they each have their strengths and weaknesses. SBOMs provide a more detailed overview of a project's component dependencies, but they do not report information on vulnerabilities. Instead, they can be used to search for known vulnerabilities in the databases using the project's components. SCA tools typically provide detailed vulnerability information but provide less information on the component dependencies themselves, and on dependency visualization. 
Their main goal is to list vulnerable components, not to expose how each one is connected to the root project. Their outputs are typically flat, i.e. for each component, they list known vulnerabilities but do not list the dependency leading to that vulnerability, as shown in Figure~\ref{dependency-chain-image}(b).

The separation between tools prevents developers from accessing the entire view of their project's component dependencies and resulting vulnerability information in a consistent manner. Figure~\ref{dependency-chain-image} illustrates this challenge visually. Using only an SBOM, a developer sees detailed dependency structure but lacks vulnerability context (see Figure~\ref{dependency-chain-image}(a)). Conversely, an SCA tool's output might flag a vulnerable component but obscure the potentially complex chain of intermediate dependencies connecting it back to the main project (see Figure~\ref{dependency-chain-image}(b)).

There are only a few tools that consolidate and unify these two data sources. Some SCA tools such as OSV-Scanner~\cite{osv} or Dependency-Track~\cite{dependency-track} allow the user to input an SBOM as a reference file; however, even when paired together, the tools are not well-suited for developers to interoperate/access complex data. Doing so often requires significant tedious manual correlation between the outputs of separate tools. Consider the specific challenge, depicted conceptually in Figure~\ref{dependency-chain-image}, where a developer identifies a vulnerability associated with a transitive dependency (like the component \texttt{flink-parent} depending on \texttt{jettison} via \texttt{flink-runtime}, \texttt{hadoop-common}, and \texttt{jersey-json}). Determining the exact dependency chain(s) using only the raw SBOM requires tedious parsing; on the other hand, the SCA tool alone might not reveal this full path. Furthermore, multiple paths might exist, complicating a manual analysis of how the root component is exposed. Relying solely on the separate outputs of the SBOM and SCA tools makes gaining this complete understanding difficult for complex projects.

To address this challenge in the software development life-cycle (SDLC), we propose an SBOM-SCA \emph{knowledge graph} (KG), where the data from both tools are reconciled and consolidated into a graph data structure. We name this graph methodology \vdgraph\  (which stands for vulnerability-dependency graph). \vdgraph\ provides a clear and holistic view of a project's open-source vulnerabilities, where developers can easily access complex vulnerability-component relationships and properties. As visualized in Figure~\ref{dependency-chain-image}(c), the KG integrates the full dependency chain with the associated vulnerability information, resolving the fragmentation illustrated in the separate tool views.

The core of \vdgraph\ is converting and consolidating SBOM and SCA-tool outputs into a graph, where vertices represent components or vulnerabilities, and edges capture relationships such as dependencies or known vulnerability links. Our proposed solution first builds a graph from an SBOM's comprehensive list of components and their inter-dependencies with each other. To then incorporate vulnerability data from an SCA-tool, we need to identify and match a vulnerable component in the SCA-tool output to a component in the SBOM graph, and then link its resulting vulnerabilities to that component in the graph.

A key challenge that arises is pulling insights from the data across the multiple tools, as we need match the component dependencies listed across them. With varying detail and naming standards across different tools (including versions in package name, dropping minor versions/patches in name, etc.), reliably matching components across the tools is not trivial. We identify the ambiguities that arise from this and their frequency. We detail our methodology and prove that it constructs a complete and reachable graph given our tooling assumptions.  

We test our methodology on a set of projects, demonstrating how to design, run, evaluate demonstration queries for more extracting complex relationships enabled by \vdgraph\ and detail new vulnerability insights gained from the queries. 

In summary, our main contributions are as follows:
\begin{itemize}
    \item We introduce a graph-theoretical framework \vdgraph\ to fuse SBOM and SCA data into a unified graph schema. We additionally characterize its space and computational complexity and theoretical properties, such as completeness and reachability.
    
    \item We provide a practical implementation of \vdgraph\ using the popular CycloneDX format (for our SBOMs) and OSV-scanner (for our SCA-tool), with end-to-end automation and benchmarking of the process.    
    \item We design and evaluate two novel queries that are applied to KGs produced by \vdgraph. 
    
    \item We evaluate \vdgraph\ on 21 Java projects (based on Balliu et al.~\cite{maven-31}). Our results highlight two important aspects of vulnerability exposure: first, it demonstrates that specific common library versions act as concentrated risk points (one instance is reachable via over 150,000 dependency paths), and second, vulnerability exposure predominately emerges via deeper transitive paths, with a disproportionately low vulnerability density at depths 0-2 hops.
        
 \end{itemize}

The rest of this paper is organized as follows.
Section~\ref{sec:relatedwork} discuses related work. Section~\ref{sec:methodology} introduces the theoretical underpinnings of \vdgraph. Section~\ref{sec:implement} details its practical implementation for popular SBOM and SCA tools. Section~\ref{sec:eval} evaluates the performance and demonstrates the usefulness of \vdgraph\ with practical queries. Section~\ref{sec:conclusion} concludes the paper.

\section{Related Work}\label{sec:relatedwork}

SBOM and SCA tools have become essential for managing software supply chain security, offering visibility into components and vulnerabilities. However, several prior works highlight key limitations in their accuracy, coverage, and integration. This section reviews recent studies on the strengths and shortcomings of both tools, as well as efforts to consolidate them.

A significant challenge for SBOMs is the lack of standardization, where different formats such as CycloneDX~\cite{maven-plugin} and SPDX~\cite{spdx} exist, leading to compatibility issues and inconsistent naming schemes across tools. Bi et al.~\cite{10.1145/3654442} conduct an empirical study analyzing SBOM-related GitHub discussions across 510  projects, highlighting the challenges in SBOM adoption in their work. The challenges include inconsistent formats, internal quality issues, and insufficient tool support. Benedetti et al.~\cite{benedetti2024} perform a security analysis on the vulnerability detection capabilities of tools using SBOM, revealing that many SBOMs introduce shortcomings in the thoroughness and accuracy in identifying vulnerabilities. They highlight that SBOM generators solely relying on static metadata (i.e., Syft and Trivy) display worse performances for vulnerability detection. O'Donoughe et al.~\cite{oDonoghue2024impacts} investigate the influence of SBOM generation on vulnerability detection via a large dataset spanning several SBOM generation tools in both CycloneDX and SPDX formats, evaluating the result on common SBOM analysis tools. They uncover high variability in vulnerability reporting attributed to SBOM generation, highlighting the importance of SBOM files and their regulation for a secure SDLC. 
Security implications of non-compliant SBOMs for vulnerability management is investigated in~\cite{xiao2025jbomaudit}. The work highlights the prevalence and severity of SBOM noncompliance issues, including thousands of SBOM files which fail to disclose direct dependencies, of which 5\% are infected with vulnerabilities. Therefore SBOM files alone cannot reliably provide security compliance.

While significant progress has been made in generating and utilizing SBOMs and SCA tools, there remains a need for solutions that can effectively scale across many projects while tracking vulnerabilities based on dependencies~\cite{mirakhorli2024landscapestudyopensource}. While tools like CycloneDX and Syft can generate SBOMs, scaling this process remains an active challenge. Pereira et al.~\cite{pereira2024automatingsbomgenerationzeroshot} propose a semantic similarity approach for automated SBOM generation, by training a transformer model to relate a product name to corresponding version strings. Sharma et al.~\cite{sharma2025understanding} review several popular SCA tools, comparing their key functionalities such as workflows, vulnerability coverage and false positive rates. They show that tools relying on the same vulnerability databases provide similar results. However, there are discrepancies on the reported vulnerabilities between SCA tools using different databases. One of the underlying issues with these discrepancies is the inability to match the package names provided by the development environment to the ones in the databases, due to different naming conventions (purl, CPE, or SWID). While it is surely convenient to create a unified naming convention, it can be foreseen that this will only introduce yet another naming convention and will not be backward-compatible without a manual effort to convert each naming convention to the 'unified' convention. Instead, we make an effort to include each recognized naming convention to provide a complete view of a graph-theoretic framework that merges detailed dependency information from SBOMs with vulnerability data from SCA tools into a unified, queryable knowledge graph.

To the best of our knowledge, no effort has been made to combine SBOMs with SCA tools to increase intelligence on the dependency paths leading to vulnerabilities in the software ecosystem. In our paper, we present a workflow to automate and combine the processes of generating and validating SBOM and of producing  SCA report analysis, resulting into a queryable knowledge graph. 

\section{Methodology}\label{sec:methodology}
\begin{figure*}[t]
\centering

\par
\centering 
\mylegenditem{red}{orange}{Root Component}\hspace{1em}
\mylegenditem{cyan}{cyan}{Dependency Component}\hspace{1em}
\mylegenditem{violet}{violet}{Vulnerability}
\par\vspace{1em}

\begin{subfigure}{0.45\textwidth}
\centering
\begin{tikzpicture} [node distance = 3cm, on grid, auto]
\node (q0) [state, shape=rectangle, style = {draw=red, fill=orange!30}] {flink-parent};
\node (q1) [state, shape=rectangle, style = {draw=cyan, fill=cyan!30},below left = of q0] {flink-kubernetes};
\node (q2) [state, shape=rectangle, style = {draw=cyan, fill=cyan!30}, below right= of q0] {flink-yarn};
\node (q3) [state, shape=rectangle, style = {draw=cyan, fill=cyan!30},below = 4cm of q0] {hadoop-common};
\node (q5) [state, shape=rectangle, style = {draw=cyan, fill=cyan!30},below right = of q3] {jackson-mapper-asl};
\node (q6) [state, shape=rectangle, style = {draw=cyan, fill=cyan!30}, below left = of q3] {protobuf-java};

\node (q8)[state, accepting, shape=rectangle, style = {draw=blue, fill=cyan!30}, below = 3.5 cm of q3]{woodstox-core};
\node (q7)[state, shape=rectangle,style = {draw = violet, fill = violet!30}, below left = of q8]{GHSA-3f7h-mf4q-vrm4};
\node (q9)[state, accepting, shape=rectangle, style = {draw=blue, fill=cyan!30}, left = of q8]{protobuf-java};
\node (q10) [state, shape=rectangle,style =  {draw = violet, fill = violet!30}, right = 4 cm of q7]{GHSA-c27h-mcmw-48hv};
\node (q11)[state, accepting, shape=rectangle, style = {draw=blue, fill=cyan!30}, right = of q8]{jackson-mapper-asl};

\path [-stealth, thick]
    (q0) edge node {depn.}(q1)
    (q0) edge node {depn.}(q2)
    (q0) edge node {depn.}(q3)
    (q1) edge node {depn.}(q3)
    (q2) edge node {depn.}(q3)
    (q3) edge node {depn.}(q5)
    (q3) edge node {depn.}(q6)
    (q8) edge [sloped] node {has\_v}(q7)
    (q9) edge [sloped] node {has\_v}(q7)
    (q11) edge [sloped] node {has\_v}(q10);

\end{tikzpicture}
\caption{Input of Algorithm~\ref{alg:matching_algo}, consisting of SBOM-connected components $G_{SBOM}$ (top) and SCA components linked to vulnerabilities $G_{SCA}$ (bottom).}
\label{fig:sbom-graph}
\end{subfigure}%
\hspace{2mm}
\begin{subfigure}{0.5\textwidth}
\centering
\begin{tikzpicture} [node distance = 3cm, on grid, auto]
\node (q0) [state, shape=rectangle, style = {draw=red, fill=orange!30}] {flink-parent};
\node (q1) [state, shape=rectangle, style = {draw=cyan, fill=cyan!30},below left = of q0] {flink-kubernetes};
\node (q2) [state, shape=rectangle, style = {draw=cyan, fill=cyan!30}, below right= of q0] {flink-yarn};
\node (q3) [state, shape=rectangle, style = {draw=cyan, fill=cyan!30},below = 4cm of q0] {hadoop-common};
\node (q5) [state, shape=rectangle, style = {draw=cyan, fill=cyan!30},below right=of q3] {jackson-mapper-asl};
\node (q6) [state, shape=rectangle, style = {draw=cyan, fill=cyan!30}, left = of q5] {protobuf-java};
\node (q4) [state, accepting, shape=rectangle, style = {draw=blue, fill=cyan!30}, left = of q3] {woodstox-core};
\node (q8) [state, shape=rectangle,style =  {draw = violet, fill = violet!30}, below = of q5]{GHSA-c27h-mcmw-48hv};
\node (q7)[state, shape=rectangle,style = {draw = violet, fill = violet!30}, left = 4cm of q8]{GHSA-3f7h-mf4q-vrm4};

\path [-stealth, thick]
    (q0) edge node {depn.}(q1)
    (q0) edge node {depn.}(q2)
    (q0) edge node {depn.}(q3)
    (q1) edge node {depn.}(q3)
    (q2) edge node {depn.}(q3)
    (q0) edge [bend right = 3cm, sloped] node {depn.}(q4)
    (q3) edge node {depn.}(q5)
    (q3) edge node {depn.}(q6)
    (q4) edge [sloped] node {has\_v}(q7)
    (q6) edge node {has\_v.}(q7)
    (q5) edge node {has\_v}(q8);

\end{tikzpicture}
\caption{Output of Algorithm~\ref{alg:matching_algo}, consisting of a combined SBOM-SCA Graph $G$  providing a path from the root to each vulnerability. Notice that the component \texttt{woodstox-core} (which does not have a match in $G_{SBOM}$) is directly connected to the root node, to ensure completeness of the graph.}
\label{fig:sca-sbom-graph}
\end{subfigure}
\caption{Example of running \vdgraph\ for the \texttt{flink} project.  \vdgraph\ provides the full path to each vulnerability, listing each intermediate dependency component.}
\end{figure*}
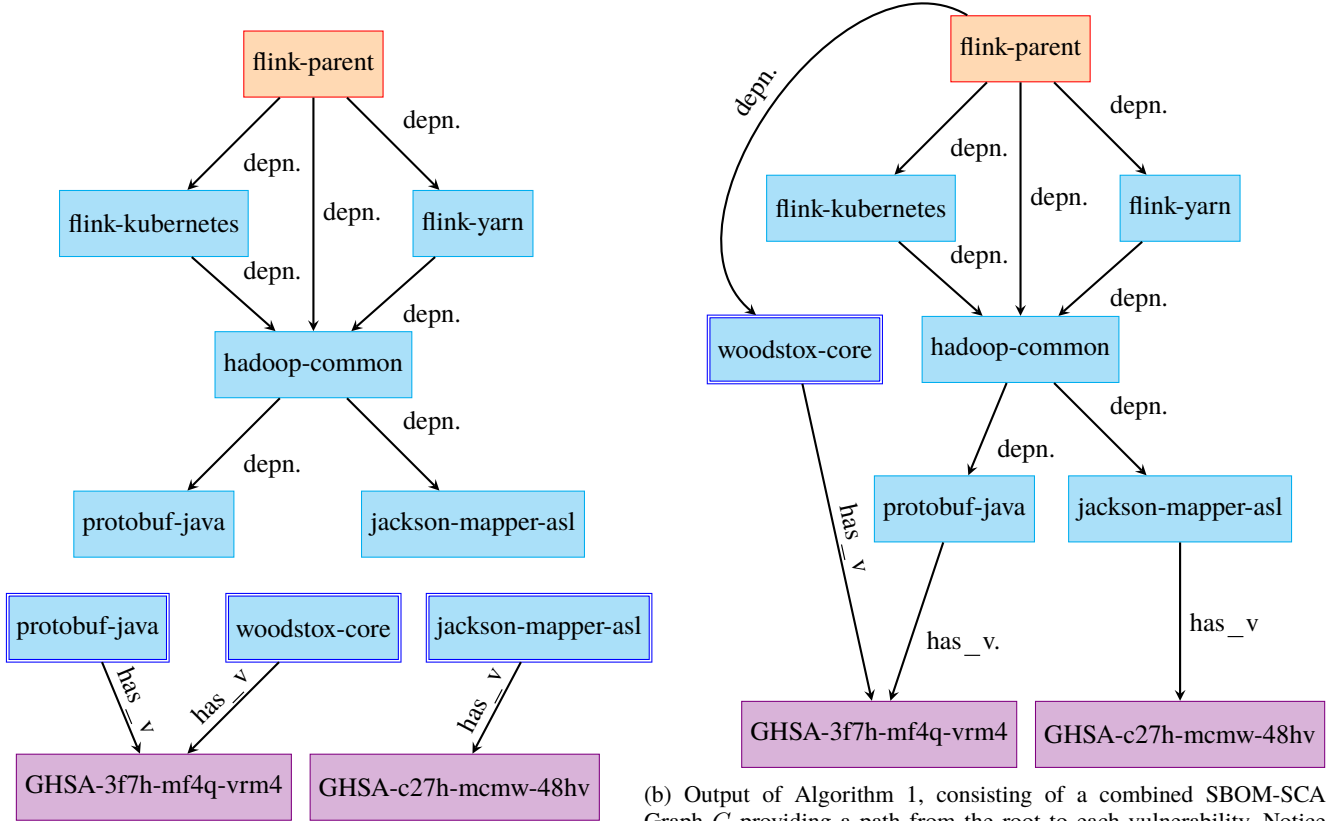

In this section, we introduce the theoretical foundations of our \vdgraph\ framework, built on top of a knowledge graph (KG) ontology. A graph only consists of vertices and edges with no other attributes. When employing a  knowledge graph, vertices and edges also have property fields represented via key-value pairs. 

\subsection{Ontology}

In this work, we use a labeled property graph (LPG) ~\cite{gallian2012graph} for representing the knowledge graph. 
More formally, we use a labeled graph notation $G(V,E,L_V,L_E)$, where $V$ represents the set of vertices, $E$ the set of (directed) edges, $L_v$ the set of vertex labels, and $L_E$ the set of edge labels. 
The root node represents the main software project under analysis, and has some key purposes such as being an anchor for the dependency graph, as it it used as a starting point from which all direct and transitive dependencies are traced. This way, by defining a single root, VDGraph can compute paths from the root to all components and vulnerabilities.

The set of vertex labels is $L_V = \{root, comp., vuln.\}$ for denoting the root node, software components and vulnerabilities, respectively. If vertex $v \in V$ has $label \in L_V$, it is denoted $v.label$. The vertices have the following additional properties following: $\{name, id, version, source \}$. For instance, $v.name = \texttt{flink}$ means that vertex $v$ is associated with the software package named \texttt{flink}.

Similarly, $L_E = \{depn, has\_v\}$ are edge labels respectively representing a dependency between software packages, or the presence of a vulnerability in a software package. If edge $e$ has $label \in L_E$ then it is accessed with $e.label$. 

\subsection{Producing SBOM and SCA Subgraphs}

The initial step in our framework is to run the SBOM and SCA tools, which separately produce two subgraphs: $G_{SBOM}$ and $G_{SCA}$.
In $G_{SBOM}$, the vertices are either the root or a component, and the only possible edge label is a dependency. 
In $G_{SCA}$, vertices can be either components or vulnerabilities, and the only edge label is $has\_v$, to associate a component to a vulnerability. Note that $G_{SCA}$ typically consists of a collection of disconnected subgraphs, where in each subgraph one or more components are connected to a vulnerability. This collection of subgraphs is called a \emph{forest} in the literature on graph theory.

This process is illustrated in Figure~\ref{fig:sbom-graph}.
\begin{enumerate}
\item  For the root and every component listed in the SBOM, we create a vertex and associate to the node properties such as its name, version, and id. Next, for any component vertex $v_1$ that depends on another component vertex $v_2$, we add an  edge $e=(v_1, v_2)$ with label $depn$ to the set of edges $E$. This produces $G_{SBOM}$, the subgraph consisting of SBOM components only. The top part of Figure~\ref{fig:sbom-graph} represents an SBOM subgraph. 

\item For each vulnerability in the SCA report, we create a vulnerability vertex with associated properties (such as id, description, severity etc.) and one or more component vertices connected to the vulnerability vertex. Each edge connecting a component vertex to a vulnerability vertex has a $has\_v$ label. The result is a subgraph $G_{SCA}$, which is a forest consisting of disconnected graphs that each connect component vertices to a vulnerability vertex. The bottom part of Figure~\ref{fig:sbom-graph} shows a SCA subgraph consisting of two smaller subgraphs.
\end{enumerate}

\subsection{VDGraph: Merging the SBOM and SCA Subgraphs} \label{sec:building_and_matching}

Algorithm~\ref{alg:matching_algo} provides a pseudo-code for the \vdgraph\ algorithm. The output of the algorithm is a graph $G$ merging $G_{SBOM}$ and $G_{SCA}$, see Figure~\ref{fig:sca-sbom-graph} for an example. Initially, $G$ is set to $G_{SBOM}$ (line 1). 
Next, for each component vertex~$u$ in $G_{SCA}$, we check if there is a matching component in $G$. As discussed in Section~\ref{sec:implement}, this step is non-trivial due to different naming conventions employed by SBOM and SCA tools. 
 There are two distinct cases:
\begin{itemize}
    \item If there is a  matching component in the dependency graph, where the names and versions both match across the existing SBOM node and the component from the SCA report, we merge the two vertices (line 5). 
    \item If no match is found, we add a new vertex to $G$ together with an edge from the root to that vertex (lines 8-9). This occurs if the SCA report lists component(s) not present in the SBOM. 
\end{itemize}
Subsequently, any vulnerability vertex $w$ connected to the component vertex $u$ is added to $G$ together with the labeled edge connecting $u$ to $w$ (lines 12-18).
Note that a vulnerable component in the SCA subgraph may match multiple components in the SBOM subgraph. In this case, the vulnerable component is separately merged with each of them. This may happen, for instance, if the SCA output does not precisely specify the software version. 

Algorithm~\ref{alg:matching_algo} has polynomial complexity. Indeed, suppose that we represent graph $G$ using an adjacency matrix, then the worst-case space complexity of Algorithm~\ref{alg:matching_algo} is $O((|V_{SBOM}|+|V_{SCA}|)^2)$, where $|V_{SBOM}|$ and $|V_{SCA}|$ represent the total number of vertices in $G_{SBOM}$ and $G_{SCA}$, respectively. This worst-case is achieved if none of the vertices in $G_{SCA}$ can be merged with vertices in $G_{SBOM}$, resulting in an adjacency matrix with $|V_{SBOM}|+|V_{SCA}|$ rows and $|V_{SBOM}|+|V_{SCA}|$ columns. Note that entries of the adjacency matrix represent the edges of $G$. Since each edge in $G$ is updated at most once (see lines 7 and 17), the worst-case computational complexity of Algorithm~\ref{alg:matching_algo} is also $O((|V_{SBOM}|+|V_{SCA}|)^2)$. As shown in Section~\ref{sec:eval}, this complexity is practical. 

\begin{algorithm}[t]
\caption{VDGraph($G_{SBOM}, G_{SCA}$)}
\label{alg:matching_algo}{}
\begin{algorithmic}[1]
\State Initialization: $G = G_{SBOM}$
\ForAll{component vertex $u \in G_{SCA}$} 
    \If{$\exists u' \in G$ such that $u.name = u'.name$ and $u.version = u'.version$} 
   \State \Comment{\textcolor{blue}{Match found}}
        \State Merge $u$ with $u'$ in $G$ 
    \Else 
    \State \Comment{\textcolor{blue}{No match found, create new vertex in $G$ and connect root to it}}
        \State Add vertex $u$  into $G$
        \State Add edge $(root, u)$ with label $depn.$ into $G$
    \EndIf   
    \State \Comment{\textcolor{blue}{Next, add vulnerability info related to $u$ into $G$}}
    \ForAll{ edge $(u, w) \in G_{SCA}$}
    
        \If {$w$ $\not \in G$}     
         \State \Comment{\textcolor{blue}{Add vulnerability vertex $w$ into $G$, if not there already}}
        \State Add vertex $w$ into $G$
        \EndIf
        \State Add edge $(u, w)$ with label $has\_v$ into $G$
        \EndFor      
    \EndFor    
\State\Return graph $G$
\end{algorithmic}
\end{algorithm}

\subsection{Properties}
We next present theoretical properties of the graph $G$.
We first prove that $G$ is complete, in the sense that it contains all the SBOM and SCA data captured by vertices and edges of the subgraphs $G_{SBOM}$ and $G_{SCA}$.

\begin{prop}
Graph $G$ is complete.
\end{prop}
\begin{proof}
Line~1 of Algorithm~\ref{alg:matching_algo} ensures that all the vertices and edges of $G_{SBOM}$ are included in $G$. Next, lines 5 and 7 ensures that each component vertex of $G_{SCA}$ is included in $G$. Next, line 15 ensures that each vulnerability vertex of $G_{SCA}$ is included in $G$. Last, line 17 ensures that each edge of $G_{SCA}$ is included in $G$.
\end{proof}
We note that there is a trade-off between completeness and accuracy. Thus, if an SCA component vertex does not match any SBOM component vertex, line~9 adds an edge between the root and the SCA component vertex. This edge may be inaccurate. On other hand, if each SCA component vertex uniquely matches an SBOM component vertex, then graph $G$ is both complete and accurate.

Next, we prove that all the vulnerabilities in $G$ are reachable from the root. This proof is based on the natural assumption that each dependency in the SBOM is reachable from the root (i.e., there exists a path between the root vertex to any component vertex in $G_{SBOM}$). 
\begin{prop}
Every vulnerability vertex in $G$ is reachable from the root vertex.
\end{prop}
\begin{proof}
Every vulnerability vertex in $G$ is connected to a component that is either merged with an SBOM component vertex (line 5) or to a component directly connected to the root (line 9). It follows that there exists a path between the root and each vulnerability vertex.
\end{proof}

\subsection{Graph Queries}\label{graph-queries}
The completeness and reachability properties of graph $G$ enable a host of queries that can offer deep insights on the scope of vulnerabilities.

At the basic level, the graph allows for direct lookup queries based on vertex labels and properties. For instance, one can query the graph for all vulnerability vertices with high severity. Such queries can be refined with additional property filters, such as finding components from a specific publisher or vulnerabilities with a certain ID.

Queries can also identify specific patterns, such as paths of a fixed length. For example, we can find all component nodes directly linked to high severity vulnerabilities (paths of length 1). Conversely, starting from a specific vulnerability vertex, one can query for all component vertices directly linked to that vertex by initiating a path traversal from the vulnerability.

More sophisticated graph algorithms, such as breadth-first search (BFS) or depth-first search (DFS), can be utilized to explore more complex patterns in the graph, such as variable length paths. 
The dependency chain example shown in Figure~\ref{dependency-chain-image} is an example of a variable length path. Finding such paths whether seeking a specific chain, the shortest path, or all paths, may require the use of an advanced traversal algorithm. Fortunately, modern tooling and graph database systems, such as Neo4j, abstract this complexity, hence allowing users to define path requirements while the query engine automatically executes optimized algorithms.


To demonstrate the usefulness of our methodology, we next describe two specific queries that we further elaborate on in Sections~\ref{sec:implement} and~\ref{sec:eval}.

\emph{\textbf{Query 1: Path Counts.}} In this query, we investigate the number of distinct paths between the root to each vulnerability vertex with high severity. 
A high path count suggests that a component has widespread reachability within the project's dependency structure. 
For each matching component, it returns the number of distinct paths from the root to the component. This query enables project developers to pinpoint and quickly assess the potential impact of vulnerable components, and allow them to focus their patching effort on the most critical components.

\emph{\textbf{Query 2: Vulnerability Depth.}} In the second query, we investigate how far and deeply nested from the root component each vulnerability is. Knowing vulnerability depth is key for prioritizing patches, as vulnerabilities closer to the core components might pose a more immediate risk or be simpler to address than those deeply nested within transitive dependencies. This depth is quantified in the graph as the shortest path between a vulnerability and the root component. BFS can be utilized to determine the minimum number of dependency hops from the root vertex to a vulnerability vertex. Our second query identifies the shortest path between the root and each vulnerability and return the length of each path.

\section{Implementation} \label{sec:implement}

This section details our implementation of \vdgraph\ that constructs a knowledge graph (KG) out of the SBOM file and SCA tool report for a software projects. We explain the processes of data acquisition, graph definition, mapping of vertices and edges, matching components, graph querying and automation.  

\subsection{Data Acquisition} \label{data_aq}
The first step is to find or generate our data sources: SBOMs and SCA tool outputs. We specifically target tools that have command line interfaces for simplified automation and a JSON-format output for easier processing. 

Several tools can build an SBOM for a software project, each with its own benefits and disadvantages~\cite{oDonoghue2024impacts}. Our implementation uses the CycloneDX Maven plugin to generate SBOMs for selected projects. The plugin is a build-time SBOM tool, creating SBOMs in the CycloneDX format during compilation. We selected it primarily because its integration with the Maven build process allows for straightforward and consistent automation of SBOM generation into either JSON or XML documents. This choice means that our implementation target Maven-based Java projects.

A sample JSON SBOM for the project \texttt{flink}\footnote{https://github.com/apache/flink.git} is shown in Listing~\ref{example-sbom}. The first major section \texttt{metadata} contains details about the document itself and the root component. The next section \texttt{components} lists out the components and their properties. The \texttt{bom-ref} field is the key identification field that differentiates different components. In our case, the value is equivalent to the components' Package URL (purl)~\cite{purl}. The final section \texttt{dependencies} lists the dependencies between components using \texttt{bom-ref} to reference components. 

Like SBOM generation, there are numerous popular SCA tools. For our implementation, we selected Google's OSV-Scanner~\cite{osv}. Its command line interface enables easy automation and its ability to change the output into a JSON format simplifies processing. OSV-Scanner identifies components by scanning user-specified inputs (like an SBOM file or directory). It then queries the OSV public vulnerability database for known vulnerabilities associated with these identified components.
In our implementation, we use the generated CycloneDX SBOM as the direct input to OSV-Scanner.

A sample JSON-formatted OSV-Scanner output for the project \texttt{flink} is shown in Listing~\ref{example-osv}. The first section \texttt{source} details what directory or file the subsequent components were found. In our case, all our components originate from the SBOM file. The next section \texttt{packages} lists the components. Each component is followed by a list of its associated vulnerabilities. Notably, the only provided fields for the components are \texttt{name}, \texttt{version}, and \texttt{ecosystem}. The \texttt{name} field is formatted with the group name appearing before the actual component name.

\begin{figure}[h]
\centering
\lstset{
  language=JSON
}
\begin{lstlisting}[caption={Example of SBOM file for the \texttt{flink} project.}, label={example-sbom}]
{
  "bomFormat" : "CycloneDX",
  "specVersion" : "1.4",
  "serialNumber" : "urn:uuid:735aa69b-bd0a...",
  "version" : 1,
  (*@"metadata"@*) : {
    "timestamp" : "2025-03-19T01:03:52Z",
    "tools":[...],
    "component" : {
      "publisher" : "The Apache Software Foundation",
      "group" : "org.apache.flink",
      "name" : "flink-parent",
      "version" : "2.1-SNAPSHOT",
      "bom-ref" : "pkg:maven/org.apache.flink/flink-parent@2.1-SNAPSHOT?type=pom", ...}
  },
  (*@"components"@*): [{
    "publisher" : "The Apache Software Foundation",
    "group" : "org.apache.flink",
    "name" : "flink-shaded-force-shading",
    "version" : "20.0", 
    "bom-ref" : "pkg:maven/org.apache.flink/flink-shaded-force-shading@20.0?type=jar", ...],
  (*@"dependencies"@*): [
  {
    "ref" : "pkg:maven/org.apache.flink/flink-parent@2.1-SNAPSHOT?type=pom",
    "dependsOn" : [
      "pkg:maven/org.apache.flink/flink-shaded-force-shading@20.0?type=jar",
      "pkg:maven/org.slf4j/slf4j-api@1.7.36?type=jar",...]
  }, ...]
}
\end{lstlisting}
\end{figure}

\begin{figure}[h]
\centering
\lstset{
  language=JSON
}
\begin{lstlisting}[caption={Example of SCA output for the \texttt{flink} project.}, label={example-osv}]
{
  "results": [
    {
      (*@"source"@*): {
      "path": "/path/to/source",
        "type": "sbom"
      },
      (*@"packages"@*): [
        {
          "package": {
            "name": "org.codehaus.jettison:jettison",
            "version": "1.1",
            "ecosystem": "Maven"
          },
          "vulnerabilities": [
          {
              "modified": "2023-11-08T04:10:53Z",
              "published": "2022-12-13T15:30:26Z",
              "schema_version": "1.6.0",
              "id": "GHSA-7rf3-mqpx-h7xg", ...},
        ...]},
...]
}    
\end{lstlisting}
\end{figure}

\subsection{Defining the Graph}
We additionally need to select a tool to represent the KG. Our implementation uses Neo4J's graph database. Neo4j is the most widely used graph database according to the DB-Engines\cite{DBEnginesGraph2025}, a database ranking site, and supports querying into the database via the Cypher language.

Neo4j is a labeled property graph database, where vertices have properties and labels. Vertices in our ontology are either roots, components or vulnerabilities. Each vertex can have any number of properties, stored as key-value fields where the value is some sort of primitive data type such as a string. Regardless, one specific field needs to be marked as the identification field. For the component nodes, unique id field is either the component name or package url, and for the vulnerability nodes, unique id field is the vulnerability id.

Edges in our ontology either represent a directed dependency relation between components or a directed vulnerability relation from a component to a vulnerability. The edges are formatted as triples with a head entity, a relationship descriptor, and a tail entity, where the unique identification field values for vertices are used to unambiguously define the entities. 

\subsection{Mapping Vertices}
To load the documents, we use Python scripts with Python's \texttt{json} library. The data is already well structured in JSON documents but some processing is still required. We create vertices for each unique component and vulnerability found. We show examples of processed components in Table~\ref{SBOM-package-table} and Table~\ref{SCA-package-table}. 

Some fields in the tool outputs contain more complex structured data, such as lists. Querying individual elements within these complex properties directly in Cypher is cumbersome compared to querying simple scalar values due to Neo4j fields only supporting primitive data types. To enhance query efficiency and simplicity, our scripts parse these complex fields and extract key information into dedicated fields. As an illustration, Table~\ref{SBOM-package-table} shows the \texttt{all licenses} field that the SBOM originally provides and the extracted most recent license value stored in its own readily accessible \texttt{most recent license} field.

For the components in OSV-Scanner's output, we separate the group from the \texttt{name} field into its own field. Since we require an explicit identification field, we create a unique ID for OSV-Scanner components by concatenating the component name and version (e.g. \texttt{commons-io\_2.2}) for vertices corresponding to OSV-Scanner components, as there is no more specific or detailed field we can generate given OSV-Scanner's outputs. As mentioned in Section~\ref{data_aq}, CycloneDX SBOMs already provide an identification field for its components in the \texttt{bom-ref} field. We follow suit and use that as the unique identification field. We process the \texttt{name} field for SBOM components the same way we process the \texttt{name} field for SCA components, to ensure consistency.

\begin{table}[htbp]
\caption{SBOM Sample Component Portion}
\begin{center}
\begin{tabular}{|>{\centering\arraybackslash}p{2cm}|>{\centering\arraybackslash}p{5cm}|}
\hline
\textbf{Field} & \textbf{Value} \\
\hline
bom-ref:ID & pkg:maven/org.apache.flink/flink-shaded-force-shading@20.0?type=jar\\
\hline
name & flink-shaded-force-shading\_20.0\\
\hline
general name & flink-shaded-force-shading\\
\hline
group & org.apache.flink\\
\hline
version & 20.0 \\
\hline
type & library\\
\hline
publisher & The Apache Software Foundation\\
\hline
all licenses & [\{'license': \{'id': 'Apache-2.0'\}\}] \\
\hline
most recent license & Apache-2.0 \\
\hline
:LABEL & package\\
\hline
... & ... \\
\hline
\end{tabular}
\label{SBOM-package-table}
\end{center}
\end{table}

\begin{table}[htbp]
\caption{OSV-Scanner Sample Component}
\begin{center}
\begin{tabular}{|>{\centering\arraybackslash}p{2cm}|>{\centering\arraybackslash}p{5cm}|}
\hline
\textbf{Field} & \textbf{Value} \\
\hline
name:ID & flink-shaded-force-shading\_20.0\\
\hline
general name & flink-shaded-force-shading\\
\hline
version & 20.0 \\
\hline
group & org.apache.flink\\
\hline
:LABEL & package\\
\hline
\end{tabular}
\label{SCA-package-table}
\end{center}
\end{table}

\subsection{Matching Components}\label{sec:componentmatch}

A key step in our framework is matching components and identifying which components are equivalent across tools. As discussed in Section~\ref{sec:building_and_matching}, we require the SBOM and the SCA-tool reports to share some sort of reliable matching key. With our generated CycloneDX SBOMS and OSV-Scanner, the most specific identification field that both tools have are the name and version. We therefore use this combined name and version to match components in our scripts. If a match is found, then we create the component vertex detailed in the SBOM (like the one shown in Table~\ref{SBOM-package-table}) as it contains more details compared to the component vertex detailed in OSV-Scanner (like the one shown in Table~\ref{SCA-package-table}). If a match is not found, only then do we create a new component vertex from OSV-Scanner. 
However, this never occurs with our framework since the OSV-Scanner's component list is sourced from the SBOM, meaning there will never be a component in the OSV-Scanner's output that is not in the SBOM.

\subsection{Mapping Edges}
 We have two types of edges, dependencies and vulnerability links. The dependency edges are sourced from the \texttt{dependencies} section in the SBOM, with the components' \texttt{bom-ref} fields being used to define the head and tail entities for the dependency edge. The vulnerability links are sourced from the OSV-Scanner output, where components are linked to their associated vulnerabilities. If a component in OSV-Scanner matches with multiple components in the SBOM, we link its associated vulnerabilities to each matching component.

\subsection{Querying}\label{sec:querying}

\begin{lstlisting}[language=cypher, caption={Cypher Implementation of Query 1 (Path Counts)}, label={cypher-ex-reachable}]
(*@MATCH@*) path=(r:root)-[:dependency*]->(c:component)
(*@WHERE@*) EXISTS {
  (*@MATCH@*) (c)-[:vulnerability]->(v:vulnerability)
  (*@WHERE@*) v.severity = 'HIGH'
}
(*@RETURN@*) c.name, count(*)
\end{lstlisting}

\begin{lstlisting}[language=cypher, caption={Cypher Implementation of Query 2 (Vulnerability Depth)}, label={cypher-ex-shortest}]
(*@MATCH@*) (r:root), (v:vulnerability) 
(*@MATCH@*) path = shortestPath((r)-[*]->(v)) 
(*@RETURN@*) length(path);
\end{lstlisting}

Once the graph is constructed in Neo4j, it can be hosted as a graph database and queried. The Cypher querying language supports various query algorithms, enabling us to query for complex patterns and relationships. Using the queries from  Section~\ref{graph-queries}, Listing~\ref{cypher-ex-reachable} demonstrates the Cypher implementation of Query 1 and Listing~\ref{cypher-ex-shortest} demonstrates the Cypher implementation of Query 2.

\subsection{Automation}
Our implementation is automated and scaled to work with multiple projects. We use scripts to automate the entire workflow: generating the SBOMs, scanning the SBOMS with OSV-Scanner, converting the documents into formats for Neo4j's KG database with the Python scripts, loading this data into the Neo4j database, and querying the database. This process can be repeated for multiple projects, allowing us to query several projects and gather vulnerability insights across them.

\section{Experimental Evaluation}\label{sec:eval}\raggedbottom
In this section, we evaluate the performance of \vdgraph\ across various projects. We then present and analyze the results obtained from running the two queries detailed in Subsections~\ref{graph-queries} and~\ref{sec:querying} to demonstrate the usefulness of our framework.  

We used \vdgraph\ to generate and query knowledge graphs for a set of projects that is based on Balliu et al.~\cite{maven-31}. We selected 21 projects that met a modified version of Balliu et al.~\cite{maven-31}'s criteria: (i)~the project uses Maven for builds; (ii)~its most recent commit was made no earlier than July 2024 (i.e., the project is actively maintained); (iii)~it declares at least one dependency in its POM file; (iv) it has received over 100 stars on GitHub. 

\subsection{Performance of VDGraph}
We detail statistics for each project in Table~\ref{project-stats}. For each project, the table lists the total number of components identified (from the generated SBOM), the number of vulnerabilities detected by the SCA tool (using OSV-Scanner), and our Python script runtimes to convert our SBOM and OSV-Scanner output into a graph representation. The scripts were ran on a MacBook Pro 14 inch 2023 Apple M3 Pro with 18~GB of RAM. Table~\ref{project-stats} confirm the practicality of \vdgraph\, as the running time does not exceed 4 seconds for any of the projects.

Our workflow uses the generated SBOM as the direct input for OSV-Scanner. Therefore, all components analyzed by OSV-Scanner are inherently also present in the SBOM data. Thus, in our experiments, \vdgraph\ never need to add an component from OSV-Scanner report that did not appear in the SBOM. Cases where multiple component entries share identical names and versions used to match SBOM and OSV-Scanner outputs, although rare, do occur. Of the 4316 total component entries aggregated from all project SBOMs, only 127 (less than 3\%) shared a name-version identifier with another component in the same project KG. This had a even smaller effect on vulnerability mapping: only a single vulnerability across all projects finding corresponded to a pair of components with identical name-version identifiers.

\begin{table}[t]
\centering
\caption{Project Statistics}
\rowcolors{2}{white}{gray!10} 
\begin{tabularx}{\linewidth}{ 
>{\raggedright\arraybackslash}p{1.8cm}
>{\raggedleft\arraybackslash}p{1.7cm} 
>{\raggedleft\arraybackslash}p{1.8cm} 
>{\raggedleft\arraybackslash}p{1.5cm}
}
\toprule
\textbf{Project Name} & \textbf{Components} & \textbf{Vulnerabilities} &
\textbf{Runtime (seconds)} \\
\midrule
accumulo & 174 & 5 & 0.797 \\
alluxio & 754 & 110 & 3.665 \\
async-http-client & 34 & 0 & 0.647 \\
checkstyle & 37 & 0 & 0.647 \\
commons-configuration & 24 & 1 & 0.658 \\
CoreNLP & 31 & 0 & 0.670  \\
error-prone & 45 & 0 & 0.644  \\
flink & 685 & 59 & 2.236 \\
handlebars.java & 81 & 6 & 0.689  \\
javaparser & 17 & 0 & 0.734 \\
jenkins & 99 & 1 & 0.700 \\
jooby & 494 & 3 & 0.928 \\
launch4j-maven-plugin & 44 & 0 & 0.633 \\
mybatis-3 & 9 & 0 & 0.635 \\
neo4j & 447 & 5 & 0.961 \\
orika & 15 & 2 & 0.639  \\
para & 233 & 3 & 2.421 \\
pitest & 140 & 8 & 0.700 \\
spoon & 17 & 0 & 0.627 \\
undertow & 29 & 1 & 0.666 \\
zerocode & 120 & 41 & 2.244 \\
\bottomrule
\end{tabularx}
\label{project-stats}
\end{table}

\subsection{Evaluation of Query 1: Path Counts to Vulnerable Components of High Severity}

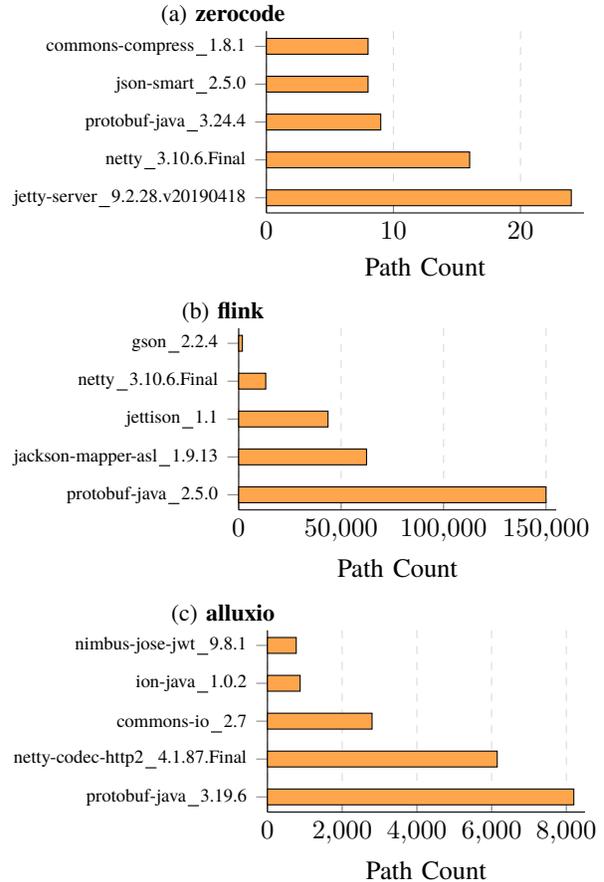
\begin{figure}[t]

  \begin{subfigure}[b]{0.32\textwidth}
    \centering
    \caption{\textbf{zerocode}}
    \begin{tikzpicture}
    \pgfplotsset{height= 4cm}
      \begin{axis}[
        xbar,
        width=\linewidth,
        bar width=6pt,
        xmin=0,xmax=25,
        xlabel={Path Count},
        symbolic y coords={
          commons-compress\_1.8.1,
          json-smart\_2.5.0,
          protobuf-java\_3.24.4,
          netty\_3.10.6.Final,
          jetty-server\_9.2.28.v20190418},
        ytick=data,
        y dir=reverse,
        y tick label style={font=\scriptsize},
        xmajorgrids,
        grid style={dashed,gray!30},
        axis lines*=left,
      ]
        \addplot+[fill=orange!70,draw=black]
          coordinates {
            (8 ,commons-compress\_1.8.1)
            (8 ,json-smart\_2.5.0)
            (9 ,protobuf-java\_3.24.4)
            (16,netty\_3.10.6.Final)
            (24,jetty-server\_9.2.28.v20190418)
          };
      \end{axis}
    \end{tikzpicture}
    
  \end{subfigure}
  \hfill
  \begin{subfigure}[b]{0.32\textwidth}
    \centering
    \caption{\textbf{flink}}
    \begin{tikzpicture}
    \pgfplotsset{height= 4cm}
      \begin{axis}[
        xbar,
        width=\linewidth,
        bar width=6pt,
        xmin=0,xmax=155000,
        xlabel={Path Count},
          x tick label style={/pgf/number format/fixed},
scaled x ticks=false,
        symbolic y coords={
          gson\_2.2.4,
          netty\_3.10.6.Final,
          jettison\_1.1,
          jackson-mapper-asl\_1.9.13,
          protobuf-java\_2.5.0},
        ytick=data,
        y dir=reverse,
        y tick label style={font=\scriptsize},
        xmajorgrids,
        grid style={dashed,gray!30},
        axis lines*=left,
      ]
        \addplot+[fill=orange!70,draw=black]
          coordinates {
            (1800  ,gson\_2.2.4)
            (13200 ,netty\_3.10.6.Final)
            (43600 ,jettison\_1.1)
            (62400 ,jackson-mapper-asl\_1.9.13)
            (150000,protobuf-java\_2.5.0)
          };
      \end{axis}
    \end{tikzpicture}
    
  \end{subfigure}
  \hfill
  \begin{subfigure}[b]{0.32\textwidth}
    \centering
    \caption{\textbf{alluxio}}
    \begin{tikzpicture}
    \pgfplotsset{height= 4cm}
      \begin{axis}[
        xbar,
        width=\linewidth,
        bar width=6pt,
        xmin=0,xmax=8500,
        xlabel={Path Count},
        symbolic y coords={
          nimbus-jose-jwt\_9.8.1,
          ion-java\_1.0.2,
          commons-io\_2.7,
          netty-codec-http2\_4.1.87.Final,
          protobuf-java\_3.19.6},
        ytick=data,
        y dir=reverse,
        y tick label style={font=\scriptsize},
        xmajorgrids,
        grid style={dashed,gray!30},
        axis lines*=left,
      ]
        \addplot+[fill=orange!70,draw=black]
          coordinates {
            (770 ,nimbus-jose-jwt\_9.8.1)
            (875 ,ion-java\_1.0.2)
            (2800,commons-io\_2.7)
            (6150,netty-codec-http2\_4.1.87.Final)
            (8200,protobuf-java\_3.19.6)
          };
      \end{axis}
    \end{tikzpicture}
    
  \end{subfigure}

    \caption{Top vulnerable components (with high severity) for three projects: \texttt{zerocode}, \texttt{flink}, and \texttt{alluxio}. The path count represents the number of distinct paths from the root node to the vulnerable component. As shown, some vulnerable components are reachable through a very large number of different paths.}
    \label{fig:vulns}
\end{figure}

We first ran Query~1 (Listing~\ref{cypher-ex-reachable}) on the projects to identify components that are specifically impacted by high severity issues. For this analysis, we queried all components reachable from a root component and identified those with direct links to high severity vulnerabilities. We then return the components that matched this criteria and together with the number of distinct paths from the root component to these components. 

Figure~\ref{fig:vulns} reports the most reachable components linked to high-severity vulnerabilities across three example projects: \texttt{zerocode}, \texttt{flink}, and \texttt{alluxio}. For each project, the x-axis represents the number of distinct dependency paths from the root to a vulnerable component, and the y-axis lists the specific component and version. This query identifies components that are not only vulnerable but are also extensively reachable within their dependency graphs. Notably, in \texttt{flink}, the vulnerable component \texttt{protobuf-java\_2.5.0} is reachable via over 150,000 paths. 
Similarly, \texttt{jetty-server\_9.2.28.v20190418} in \texttt{zerocode} and \texttt{protobuf-java\_3.19.6} in \texttt{alluxio} are also highly reachable, emphasizing their criticality in transitive dependency chains.

These findings reinforce the importance of identifying not just whether a project is vulnerable, but \textit{how} and \textit{where} those vulnerabilities are introduced. Even if a root project does not directly depend on a vulnerable component, high severity risks propagate through commonly shared or transitively included packages as seen in our example. 

\subsection{Evaluation of Query 2: Vulnerability Depth}

\pgfplotstableread[row sep=\\,col sep=&]{
        Label & Low & Med & High & Crit \\
        1 & 0 & 0 & 0 & 0 \\
        2 & 5 & 11 & 9 & 1 \\
        3 & 8 & 38 & 25 & 8 \\
        4 & 5 & 36 & 26 & 7 \\
        5 & 1 & 10 & 9 & 3 \\ 
        6 & 0 & 9 & 4 & 2 \\ 
        7 & 1 & 2 & 4 & 0 \\
        8 & 0 & 7 & 6 & 3 \\
        9 & 0 & 0 & 2 & 0 \\ }\shortestpathdata
    
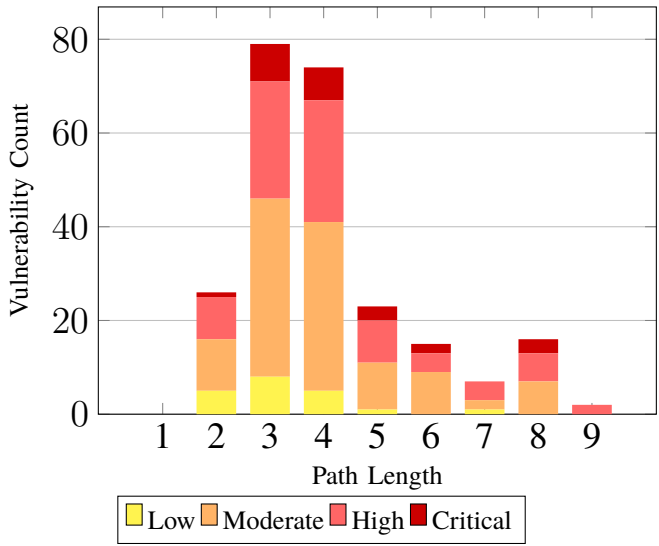
\begin{figure}
    \centering
    \begin{tikzpicture}
    \pgfplotsset{width=9cm, height= 7cm, compat=1.9,compat/bar nodes=1.8}
    \pgfplotsset{every tick label/.append style={font=\Large}}
    \begin{axis}[
        ymin=0,
        ybar stacked,
        ymajorgrids,
        bar width=15pt,
        enlarge x limits=0.15,
        legend style={at={(0.4,-0.20)},
        anchor=north,legend columns=-1},
        ylabel={Vulnerability Count},
        xlabel={Path Length},
        xticklabels from table={\shortestpathdata}{Label},
        xtick=data,
    ]
    \addplot [fill=yellow!80, draw=none] 
        table [y=Low, meta=Label, x expr=\coordindex]
            {\shortestpathdata};\addlegendentry{low}
    \addplot [fill=orange!60, draw=none] 
        table [y=Med, meta=Label, x expr=\coordindex]
            {\shortestpathdata};\addlegendentry{med}
    \addplot [fill=red!60, draw=none] 
        table [y=High, meta=Label, x expr=\coordindex]
            {\shortestpathdata};\addlegendentry{high}
    \addplot [fill=red!80!black, draw=none] 
        table [y=Crit, meta=Label, x expr=\coordindex]
            {\shortestpathdata};\addlegendentry{critical}   
    \legend{\strut Low, \strut Moderate, \strut High, \strut Critical}
    \end{axis}
    \end{tikzpicture}
    \caption{Distribution of the length of the shortest paths from the root to each vulnerable component, for all projects listed in Table~\ref{project-stats}. Different colors indicate different vulnerability severity. The figure shows that most vulnerable components are three or four hops away from the root.}
    \label{all_shortest-path-freq}
\end{figure}

Figure~\ref{all_shortest-path-freq} shows the results from running Query 2 (Listing~\ref{cypher-ex-shortest}) on each of the projects. We tracked the shortest dependency paths from the root component to each vulnerability across all projects. The mean number of intermediate dependencies is 4.07  and the median is 4.  We find that 74.0\% of all vulnerabilities are within four or fewer component dependencies away from the root vertex, and likewise 69.7\% of critical or high severity vulnerabilities. Yet, there is no vulnerable component that is one-hop away from the root and relatively few vulnerable components that are two-hop away.   

\begin{figure}
    \centering
    \begin{tikzpicture}
    \pgfplotsset{width=9cm, height= 7cm, compat=1.9,compat/bar nodes=1.8}
    \pgfplotsset{every tick label/.append style={font=\Large}}
    \begin{axis}[
        ymin=0,
        ybar stacked,
        ymajorgrids,
        bar width=15pt,
        enlarge x limits=0.15,
        legend style={at={(0.4,-0.20)},
        anchor=north,legend columns=-1},
        ylabel={Component Count},
        xlabel={Path Length},
        xticklabels from table={\componentpathdata}{Label},
        xtick=data,
    ]
    \addplot table [y=Path Length, x expr=\coordindex] {\componentpathdata};
    \end{axis}
    \end{tikzpicture}
    \caption{Distribution of the length of the shortest paths from the root to each component (vulnerable or not), for all projects listed in Table~\ref{project-stats}. Combined with Figure~\ref{all_shortest-path-freq}, this shows that the vulnerabilities are all transitive (i.e., no directly connected component is vulnerable).}
    \label{fig:component-shortest-path}
\end{figure}
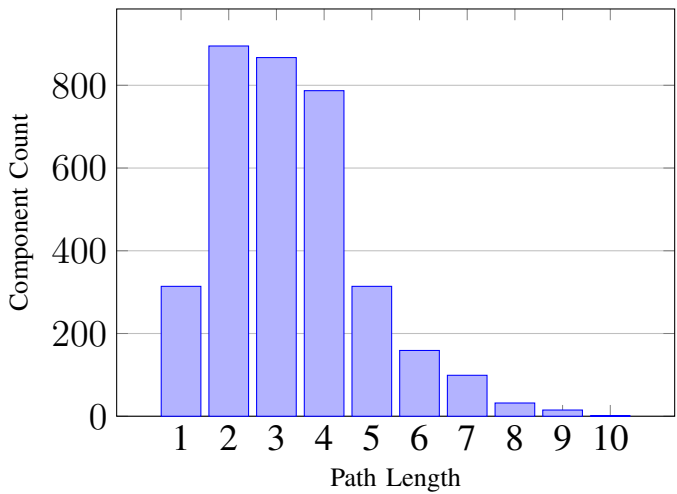

This result can be contextualized by comparing it to the depth distribution of \emph{all} components (i.e., both vulnerable and non-vulnerable components).  Figure~\ref{fig:component-shortest-path} shows that 82.2\% of the components are within four hops from the root. However, the component distribution does not align with Figure~\ref{all_shortest-path-freq}. Indeed, a significant portion (34.7\%) of all components are direct or secondary components. Proportionally, these components have much fewer vulnerabilities than those that are three-hop or four-hop away. We hypothesize that this difference arises because developers have more direct control over components that are close to the core build, allowing them to unitize more secure components and more recent versions of the components. Vulnerabilities only begin appearing in the deeper, less visible transitive dependency layers where control is diminished. This emphasizes the importance of scanning for deeply nested vulnerabilities, even if the direct dependencies themselves appear secure.

\pgfplotstableread[row sep=\\,col sep=&]{
        Label & Path Length \\
        1 & 314 \\
        2 & 895 \\
        3 & 867 \\
        4 & 787 \\
        5 & 314 \\
        6 & 159 \\
        7 & 99 \\
        8 & 32 \\
        9 & 15 \\
        10 & 1 \\}\componentpathdata
        
\section{Conclusion}\label{sec:conclusion}

In this paper, we presented \vdgraph, a framework for integrating SBOM and SCA outputs into a unified graph ontology. We detailed our methodology for data mapping, provided theoretical grounding for the framework's properties and demonstrated a practical application of \vdgraph\ via an implementation utilizing the CycloneDX Maven plugin, OSV-Scanner, and Neo4j. \vdgraph\ allows one to formulate novel queries that provide insights about the impact of each vulnerability beyond its severity score, such as the shortest distance from the root to the vulnerability and the number of different paths from the root to the vulnerability. Our evaluation of 21 Maven-based Java projects illustrated these insights.

While \vdgraph\ provides a foundation for a holistic view of vulnerability exposure, there  are limitations such as tool chain specificity (Maven, CycloneDX, OSV-Scanner) and challenges with reconciling component identifiers across various tools. Moreover, while \vdgraph\ does provide enhanced visibility and a unique view on vulnerability exposure, it does not replace the need for expert risk analysis. 
Moreover, the \vdgraph\ framework adopts a conservative model of vulnerability propagation. Any component that depends on a vulnerable package is assumed to inherit its vulnerability. This leads to an over-approximation, as it does not account for whether the vulnerable code is actually invoked, either directly or transitively. For example, a vulnerability in a deep dependency might only affect specific APIs that are never used by the parent project. While this approach simplifies graph construction and enables broad risk visibility, it can result in false positives. However, this  limitation is inherent to software composition analysis and hence, propagates to \vdgraph\.
\balance

Future work will focus on a broader evaluation of \vdgraph\ with additional projects and tool chains, enhancing the logic of component matching, and exploring the integration of data from multiple interconnected projects into a single interconnected graph ontology to unlock even richer, complex vulnerability insights across a project set.

\section*{Acknowledgments}
This work was supported in part by the Boston University Red Hat Collaboratory (awards numbers~2024-01-RH03 and 2025-01-RH05)

\bibliographystyle{IEEEtran}
\bibliography{references}

\begin{thebibliography}{10}
\providecommand{\url}[1]{#1}
\csname url@samestyle\endcsname
\providecommand{\newblock}{\relax}
\providecommand{\bibinfo}[2]{#2}
\providecommand{\BIBentrySTDinterwordspacing}{\spaceskip=0pt\relax}
\providecommand{\BIBentryALTinterwordstretchfactor}{4}
\providecommand{\BIBentryALTinterwordspacing}{\spaceskip=\fontdimen2\font plus
\BIBentryALTinterwordstretchfactor\fontdimen3\font minus \fontdimen4\font\relax}
\providecommand{\BIBforeignlanguage}[2]{{%
\expandafter\ifx\csname l@#1\endcsname\relax
\typeout{** WARNING: IEEEtran.bst: No hyphenation pattern has been}%
\typeout{** loaded for the language `#1'. Using the pattern for}%
\typeout{** the default language instead.}%
\else
\language=\csname l@#1\endcsname
\fi
#2}}
\providecommand{\BIBdecl}{\relax}
\BIBdecl

\bibitem{cobleigh2018identifying}
A.~Cobleigh, M.~Hell, L.~Karlsson, O.~Reimer, J.~S{\"o}nnerup, and D.~Wisenhoff, ``{Identifying, prioritizing and evaluating vulnerabilities in third party code},'' in \emph{2018 IEEE 22nd International Enterprise Distributed Object Computing Workshop (EDOCW)}.\hskip 1em plus 0.5em minus 0.4em\relax IEEE, 2018, pp. 208--211.

\bibitem{gangadharan2008license}
G.~Gangadharan, S.~De~Paoli, V.~D’Andrea, and M.~Weiss, ``License compliance issues in free and open source software,'' \emph{MCIS 2008 Proceedings}, p.~2, 2008.

\bibitem{mari2003impact}
Mari and Eila, ``{The impact of maintainability on component-based software systems},'' in \emph{2003 Proceedings 29th Euromicro Conference}.\hskip 1em plus 0.5em minus 0.4em\relax IEEE, 2003, pp. 25--32.

\bibitem{mirakhorli2024landscapestudyopensource}
\BIBentryALTinterwordspacing
M.~Mirakhorli, D.~Garcia, S.~Dillon, K.~Laporte, M.~Morrison, H.~Lu, V.~Koscinski, and C.~Enoch, ``{A Landscape Study of Open Source and Proprietary Tools for Software Bill of Materials (SBOM)},'' 2024. [Online]. Available: \url{https://arxiv.org/abs/2402.11151}
\BIBentrySTDinterwordspacing

\bibitem{haddad2020open}
I.~Haddad, ``{An open guide to evaluating software composition analysis tools},'' \emph{Linux Foundation}, 2020.

\bibitem{sharma2025understanding}
P.~Sharma, Z.~Shi, S.~Simsek, D.~Starobinski, and D.~S. Medina, ``{Understanding Similarities and Differences Between Software Composition Analysis Tools},'' \emph{IEEE Security \& Privacy}, vol.~23, no.~1, pp. 53--63, 2025.

\bibitem{osv}
\BIBentryALTinterwordspacing
Google, ``Google {OSV} scanner,'' 2023. [Online]. Available: \url{https://google.github.io/osv-scanner/}
\BIBentrySTDinterwordspacing

\bibitem{dependency-track}
\BIBentryALTinterwordspacing
{The OWASP Foundation}, ``dependency-track,'' 2023. [Online]. Available: \url{https://dependencytrack.org/}
\BIBentrySTDinterwordspacing

\bibitem{maven-31}
\BIBentryALTinterwordspacing
C.~Soto-Valero, N.~Harrand, M.~Monperrus, and B.~Baudry, ``{A comprehensive study of bloated dependencies in the Maven ecosystem},'' \emph{Empirical Software Engineering}, vol.~26, no.~3, Mar. 2021. [Online]. Available: \url{http://dx.doi.org/10.1007/s10664-020-09914-8}
\BIBentrySTDinterwordspacing

\bibitem{maven-plugin}
\BIBentryALTinterwordspacing
{The OWASP Foundation}, ``Cyclonedx maven plugin,'' 2024. [Online]. Available: \url{https://github.com/CycloneDX/cyclonedx-maven-plugin}
\BIBentrySTDinterwordspacing

\bibitem{spdx}
\BIBentryALTinterwordspacing
{Linux Foundation}, ``The system package data exchange™ (spdx®),'' 2021. [Online]. Available: \url{https://spdx.dev}
\BIBentrySTDinterwordspacing

\bibitem{10.1145/3654442}
\BIBentryALTinterwordspacing
T.~Bi, B.~Xia, Z.~Xing, Q.~Lu, and L.~Zhu, ``{On the Way to SBOMs: Investigating Design Issues and Solutions in Practice},'' \emph{ACM Trans. Softw. Eng. Methodol.}, vol.~33, no.~6, Jun. 2024. [Online]. Available: \url{https://doi.org/10.1145/3654442}
\BIBentrySTDinterwordspacing

\bibitem{benedetti2024}
\BIBentryALTinterwordspacing
G.~Benedetti, S.~Cofano, A.~Brighente, and M.~Conti, ``{The Impact of SBOM Generators on Vulnerability Assessment in Python: A Comparison and a Novel Approach},'' 2024. [Online]. Available: \url{https://arxiv.org/abs/2409.06390}
\BIBentrySTDinterwordspacing

\bibitem{oDonoghue2024impacts}
\BIBentryALTinterwordspacing
E.~O'Donoghue, B.~Boles, C.~Izurieta, and A.~M. Reinhold, ``{Impacts of Software Bill of Materials (SBOM) Generation on Vulnerability Detection},'' in \emph{Proceedings of the 2024 Workshop on Software Supply Chain Offensive Research and Ecosystem Defenses}, ser. SCORED '24.\hskip 1em plus 0.5em minus 0.4em\relax New York, NY, USA: Association for Computing Machinery, 2024, p. 67–76. [Online]. Available: \url{https://doi.org/10.1145/3689944.3696164}
\BIBentrySTDinterwordspacing

\bibitem{xiao2025jbomaudit}
Y.~Xiao, D.~Kirat, D.~L. Schales, J.~Jang, L.~Xing, and X.~Liao, ``Jbomaudit: Assessing the landscape, compliance, and security implications of java sboms,'' in \emph{ISOC Network and Distributed System Security Symposium}, 2025.

\bibitem{pereira2024automatingsbomgenerationzeroshot}
\BIBentryALTinterwordspacing
D.~Pereira, C.~Molloy, S.~Acharya, and S.~H.~H. Ding, ``{Automating SBOM Generation with Zero-Shot Semantic Similarity},'' 2024. [Online]. Available: \url{https://arxiv.org/abs/2403.08799}
\BIBentrySTDinterwordspacing

\bibitem{gallian2012graph}
J.~A. Gallian, ``Graph labeling,'' \emph{The electronic journal of combinatorics}, pp. DS6--Nov, 2012.

\bibitem{purl}
S.~Schuberth, P.~Ombredanne, J.~Kowalleck, W.~Bartholomew, S.~Springett, and J.~M. Horan, ``Package-url,'' \url{https://github.com/package-url/purl-spec}, 2017.

\bibitem{DBEnginesGraph2025}
\BIBentryALTinterwordspacing
DB-Engines, ``Db-engines ranking of graph dbms,'' 2025. [Online]. Available: \url{https://db-engines.com/en/ranking/graph+dbms}
\BIBentrySTDinterwordspacing

\end{thebibliography}

\end{document}